%% file: tmp.tex
\documentstyle[12pt,fleqn,epsf]{article}  % worked

\input mathmacros
\def\singlespace{\baselineskip=12pt}      % spacing for stuff like abstract
\def\author#1 {\medskip\centerline{\it #1}\smallskip}
\def\address#1{\centerline{\it #1}\smallskip}

\def\email#1{{\it \qquad\qquad internet email address: #1}}
\def\reference{\hangindent=1pc\hangafter=1} % to put before each reference
\def\ReferencesBegin
{
 \singlespace					   % single spacing
 \vskip 0.5truein
 \centerline           {\bf References}
 \par\nobreak
 \medskip
 \noindent
 \parindent=2pt
 \parskip=6pt			% earlier was 10 pt and then 4pt
 }
\font\titlefont=cmb10 scaled\magstep2 
\def\journaldata#1#2#3#4{{\it #1} {\bf #2:} #3 (#4)}
\def\eprint#1{$\langle$#1\hbox{$\rangle$}}
\def\linebreak{\hfil\break}

\begin{document}

\rightline{gr-qc/0302009}
\rightline{SU--GP--2002/12--1}

\vskip 0.3 true in

\centerline{{\titlefont Black Hole Entropy as Causal Links }\footnote%
{To appear in a special issue of {\it Foundations of Physics} in honor
 of Jacob Bekenstein, ``Thirty years of black hole physics'', edited by
 L.~Horwitz.}}

\bigskip\bigskip

\author{Djamel Dou}
\address
  {Institute of Exact Science and Technology, 
   University Center of Eloued,
   Eloued, Algeria}
\email{djsdou@yahoo.com}

\medskip

\author {Rafael D. Sorkin}
\address 
 {Department of Physics, 
  Syracuse University, 
  Syracuse, NY 13244-1130, 
  U.S.A.}
\email{sorkin@physics.syr.edu}

\begin{abstract}
We model a black hole spacetime as a causal set and count, with a
certain definition, the number of causal links crossing the horizon in
proximity to a spacelike or null hypersurface $\Sigma$.  We find that
this number is proportional to the horizon's area on $\Sigma $, thus
supporting the interpretation of the links as the ``horizon atoms'' that
account for its entropy.  The cases studied include not only equilibrium
black holes but ones far from equilibrium.
\end{abstract}

\section{Introduction}

Despite all the evidence for an entropy associated with the horizon of a
black hole, a full understanding of its statistical origin is still
lacking and it remains uncertain what ``degrees of freedom'' the entropy
refers to.\footnote%
{The question is nicely posed in [1].}
Ideally one would appeal for the answer to some more fundamental quantum
theory of spacetime structure, but unfortunately no approach to
constructing such a ``quantum gravity'' theory has advanced far enough
to offer a definitive account of what the horizon ``degrees of freedom''
might be.  Nevertheless, it is hard to doubt that black hole
thermodynamics has opened up a path leading to a better knowledge of the
small scale structure of spacetime.  Indeed, the role being played by
black hole thermodynamics in this connection looks more and more
analogous to the role played historically by the thermodynamics of a box
of gas in revealing the underlying atomicity and quantum nature of
everyday matter and radiation.  We can bring out this analogy more
clearly by recalling some facts about thermodynamics in the presence of
event horizons.

One often thinks of entropy as measure of missing or ``unavailable''
information about a physical system, and from this point of view, one
would have to expect some amount of entropy to accompany an event
horizon, since it is by definition an information hider {\it par excellence}.
In particular, one can associate to each quantum field in the presence
of a horizon the ``entanglement entropy'' that necessarily results from
tracing out the interior (and therefore inaccessible) modes of the
field, given that these modes are necessarily correlated with the
exterior modes.  In the continuum, this entanglement entropy turns out
to be infinite, at least when calculated for a free field on a fixed,
background spacetime.  However, if one imposes a short distance cutoff
on the field degrees of freedom, one obtains instead a finite entropy; and
if the cutoff is chosen around the Planck length then this entropy has
the same order of magnitude as that of the horizon.  Based on this
appealing result, there have been many speculations attributing the
black hole entropy to the sum of all the entanglement entropies of the
fields in nature.

Whether or not the entanglement of quantum fields furnishes all of the
entropy or only a portion of it, contributions of this type must be
present, and any consistent theory must provide for them in its
thermodynamic accounting.  The case appears to be similar to that of an
ordinary box of gas, where we know that, fundamentally, the finiteness
of the entropy rests on the finiteness of the number of molecules, and
to lesser extent on the discreteness of their quantum states.  Indeed,
at temperatures high enough to avoid quantum degeneracy, the entropy is,
up to a logarithmic factor, merely the number of molecules composing the
gas.  The similarity with the black hole becomes evident when we
remember that the picture of the horizon as composed of discrete
constituents gives a good account of the entropy if we suppose that each
such constituent occupies roughly one unit of Planck area and carries
roughly one bit of entropy.  A proper statistical derivation along these
lines would require a knowledge of the dynamics of these constituents,
of course.  However, in analogy with the gas, one may still anticipate
that the horizon entropy can be estimated by counting suitable discrete
structures, analogs of the gas molecules, without referring directly to
their dynamics.

Clearly, this type of estimation can succeed only if well defined,
discrete entities can be identified which are available to be counted.
Within a continuum theory, it is hard to think of such entities.
Indeed, if one accepts the estimates carried out below, the entropy
would come out infinite were spacetime a true continuum.  It would
diverge with the cutoff at the same rate as the aforementioned entropy
of entanglement of an ambient quantum field.  In causal set theory, on
the other hand, the elements of the causal set serve as ``spacetime
atoms'', and one can ask whether these elements, or some related
structures, are suited to play the role of ``horizon molecules''.
%
%% (RDS) ref [R::horizon-atoms] removed pending finding the reference
%
In this paper, we will identify a certain kind of ``causal link'' as one
such structure and we will show that the black hole entropy can be
equated to the number of such links crossing the horizon $H$ in
proximity to the hypersurface $\Sigma$ for which the entropy is sought.
Moreover, almost all of these links will turn out to be localized very
near to $H$.  In consequence, conditions deep inside the black hole will
become irrelevant to the counting, as indeed they must do if any
interpretation of the entropy in terms of ``horizon degrees of freedom''
is to succeed.

\section{Counting Links}

Before proceeding, let us briefly review the terminology we will use.
For a fuller introduction to causal sets, see [2] and
references therein.

A {\it causal set} (or ``causet'') is a locally finite, partially
ordered set.  We use $\prec$ to represent the order relation and adopt
(in this paper)
the reflexive convention, according to which every element precedes itself:
$x{\prec}x$.  Let $C$ be a causet and let $x$ and $y$ be elements of
$C$.  The past of $x$ is the subset 
$\past(x)=\SetOf{y\in C}{y\prec x}$
and its future is 
$\future(x)=\SetOf{y\in C}{x\prec y}$.
If $x,y{\in}C$, $x\prec y$, and 
$\future(x)\cap\past(y)=\braces{x,y}$ 
then we call the relation $x\prec{y}$ a {\it link}.
Note that (thanks to the local finiteness) if the links of a causet are
given, then all the other relations are implied by transitivity; hence
the whole structure of the causet is encoded in its irreducible
relations or links.
An element of a causet (or of a subcauset) is {\it maximal} 
(resp. {\it minimal}) iff it is to the past (resp. future) of no other
element in the causet (or subcauset).

Now the basic hypothesis of causal set theory is that spacetime,
ultimately, is discrete, and its deep structure is that of a partial
order rather than a differentiable manifold.  The macroscopic spacetime
continuum of 
experience must then be recovered as an approximation to the causet.
Although a more sophisticated notion 
of approximation 
might ultimately be needed
[3], the intuitive idea at work here is that of a
``faithful embedding''.  If $M$ is a Lorentzian manifold and $C$ a
causal set, then a faithful embedding of $C$ into $M$ is an 
injection
$f:C\to M$ of the causet into the manifold that satisfies the
following requirements: 
(1) The causal relations induced by the embedding agree with those of
$C$ itself, i.e  
$f(x)\in J^{-}(f(y))$ iff $x\prec y$, where
$J^{-}(p)$ stands for the causal past of $p$ in $M$;
(2) The embedded points are distributed with unit density, 
and
(3) the characteristic length over which the geometry
varies appreciably is everywhere much greater than the mean spacing
between the embedded points.  
When these conditions are satisfied,
the spacetime $M$ is said to be a
continuum approximation to $C$.  
{}From the point of view of an $M$, the causet resembles a ``random
lattice'' obtained by ``sprinkling in points'' until the required
density is reached.  Thus, the probability that there will be $n$
embedded points in a given volume $V$ is given by the Poisson
distribution, $(\varrho_c V)^{n}e^{-\varrho_c V}/n!$, where the fundamental
density $\varrho_c$ is unknown but presumed to be of Planckian magnitude.

Let us now consider the entropy associated with a horizon in a spacetime
$M$ in which a causet $C$ is faithfully embedded.  As discussed in the
introduction, we expect that the entropy can be understood as
entanglement in a sufficiently generalized sense, and we may hope to
estimate its leading behavior by counting suitable
discrete structures that measure the potential entanglement in some
way.  At the same time, we know that the entropy essentially just
measures the horizon area, whence, phenomenologically, our discrete
structures must turn out to be equal in number to the horizon area, up
to small fluctuations.\footnote%
{In fact, it seems far from obvious that such structures must exist.  If
 they do, then they provide a relatively simple, order theoretic measure
 of the area of a cross section of a null surface, and, unlike what
 one's Euclidean intuition might suggest, it is known that such measures
 are not easy to come by.  For example, no one knows such a measure of
 spacelike distance between two sprinkled points that works in even such
 a comparatively simple case as a sprinkling of Minkowski
 spacetime [4].}  
{}From both points of view, a natural candidate for the structure we seek
is a {\it link} of the causet.  Indeed, we may think heuristically of
``information flowing along links'' and producing entanglement when it
flows across the horizon during the course of the causet's growth (or
``time development'').  Since links are irreducible causal relations (in
some sense the building blocks of the causet), it seems natural that by
counting links between elements that lie outside the horizon and
elements that lie inside, one would measure the degree of entanglement
between the two regions.  Equally, it seems natural that the number of
such causal links might turn out to be proportional to the horizon area,
as desired.

\subsection{An equilibrium black hole}

Let us now consider a spherically symmetric collapsing star which
produces a black hole with horizon $H$, and let $\Sigma$ be a (null or
spacelike) hypersurface on which we wish to evaluate the horizon
entropy.  For simplicity we shall ignore the influence of the collapse
and treat the black hole metric as exactly Schwarzschild in the region
of interest.  If our picture is consistent, doing so cannot change
anything, and we will see evidence for this further into the calculation.
Thus, we will work with an eternal black hole spacetime $M$, as shown in
Figure 1.  Notice, however, that only the portion of the extended
Schwarzschild spacetime that could have arisen from a collapse is to be
taken into consideration (i.e. the region exhibited in the diagram).

% This is figure 1
\begin{figure}[t]
\epsfxsize=7truein
\centerline{\epsfbox{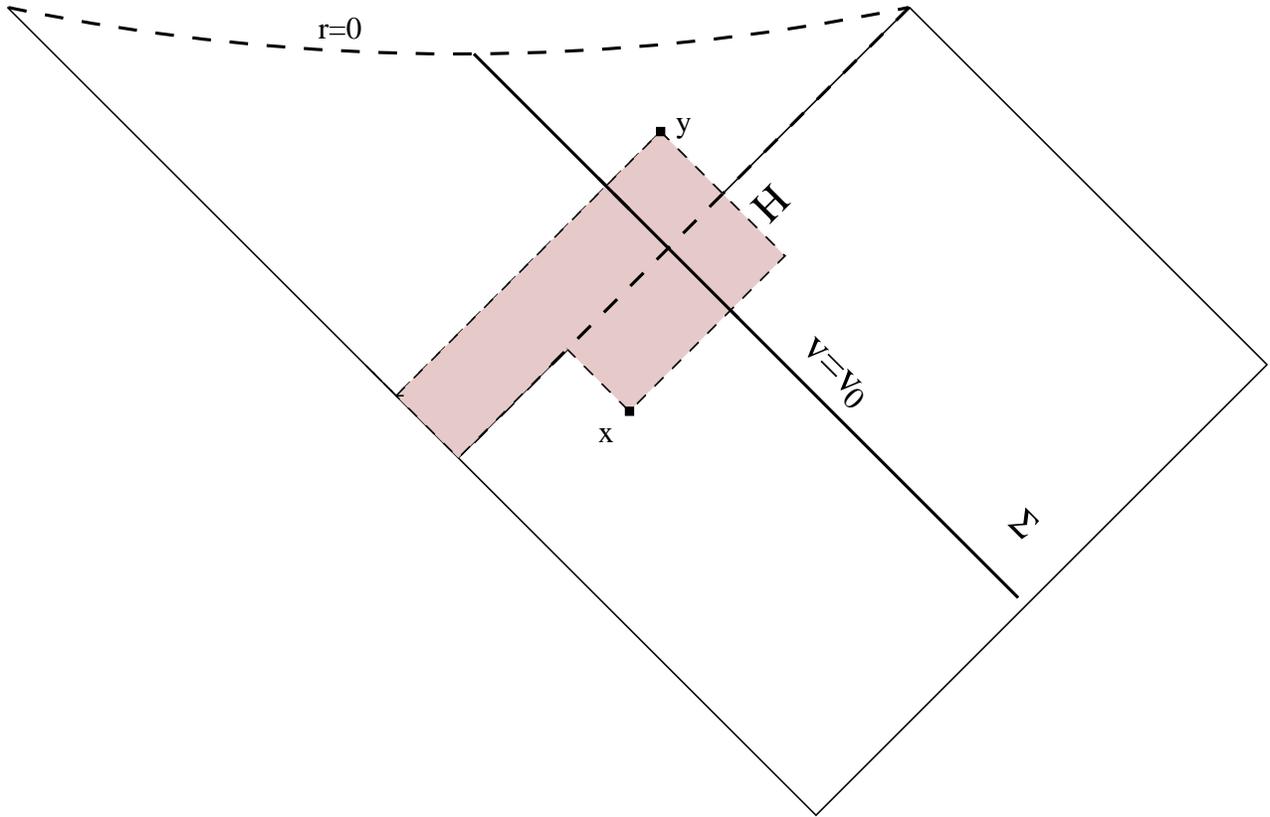}}
\caption{An equilibrium black hole and null hypersurface $\Sigma$}
\end{figure}

Now let $C$ be a causet produced by randomly sprinkling points into $M$
with density $\varrho_c=1$ in causal set units; by definition, then, $C$
is faithfully embedded in $M$.  Let $x$ be a sprinkled point in the
region $J^{-}(H)\cap J^{-}(\Sigma)$, and let $y$ be a second sprinkled
point in $J^{+}(H)\cap J^{+}(\Sigma)$.  (In other words, $x$ is outside
the black hole and to the past of $\Sigma$, while $y$ is inside the
black hole and to the future of $\Sigma$.)  To say that $x\prec{y}$ is a
link of $C$ means that the ``Alexandrov interval'',
$J(x,y):=J^{+}(x)\cap{}J^{-}(y)$, is empty of sprinkled points except
for $x$ and $y$: no sprinkled point lies causally between $x$ and $y$.
Such a pair $(x,y)$ might seem to be a good candidate for the sort of
``horizon molecule'' we wish to count.

In fact the counting reduces to the calculation of an integral, since,
as a simple consideration shows [5], the expected number
of such pairs is
$$
   \bra n \ket \; = \int_{D} e^{-V(x,y)} dV_{x} dV_{y} \ . \eqno(1)
$$
Here $V(x,y)$, whose presence serves to ensure the link condition, is
the volume of $J(x,y)$, and $D$ is the domain of integration for $x$ and
$y$.
Unfortunately, if we impose no further conditions on $x$ and $y$, then
the integral (1) can be shown to diverge when $\Sigma$ is
spacelike.
Therefore, the links we have identified cannot be the ones we want.  

To help understand the meaning of this divergence, let us remember that,
intuitively, we are trying to estimate, not the sum total of all ``lost
information'', but only that corresponding ``to a given time'', meaning
in the vicinity of the given hypersurface $\Sigma$.  Hence, to associate
one and the same causal link with more than one hypersurface would be to
``overcount'' it in forming our estimate, and it is this overcounting
that seems to be the source of our divergent answer.  Thus, what we
need is a further condition or conditions on $x$ and $y$ that would be
satisfied only by links that truly belong to $\Sigma$ rather than to
some earlier or later hypersurface.  Several possibilities suggest
themselves for this purpose, for example the requirement that $x$ be
{\it maximal} in $J^{-}(\Sigma)$, but none seems to be clearly best.
Fortunately, the end result seems to be relatively insensitive to which
choice one makes.  The precise conditions we will use will be specified
below, and the general issue will be discussed further in Section 3.

Now, ideally we would have evaluated $\bra{n}\ket$ for a fully four
dimensional Schwarzschild black hole, but unfortunately, this is
rendered difficult by the need to know all the Alexandrov neighborhoods
$J(x,y)$ of the Schwarzschild metric.  For this reason, we will simplify
the calculation by working with a ``dimensionally reduced'' two
dimensional metric instead of the true, four dimensional one.  As the
calculation proceeds, it will become very plausible that (for
macroscopic black holes) the full four-dimensional answer would differ
from the two-dimensional one only by a fixed (albeit still unknown)
proportionality coefficient of order unity, together with a factor of
the horizon area.  This will effectively accomplish our primary aim of
demonstrating that the expected number of links is proportional to the
area of the horizon in causet units.

% identifying each sphere to a point, the resulting two dimensional
% spacetime has exactly the same causal structure as the S-sector of the
% 4-dimensional one.

A radial section of a four dimensional Schwarzschild spacetime has a
line element obtained by omitting the angular coordinates from the
four dimensional line element, namely
$$
       ds^{2} = - \frac{4a^3}{r} e^{-r/a} du dv  \ ,
$$
where $a$ is the radius of the black hole (Schwarzschild radius) and $u$
and $v$ are the usual Kruskal-Szekeres coordinates, with $r$ defined
implicitly by the equation\footnote%
{Our signs are such that $u\sim t-r$, $v\sim t+r$, and the horizon $H$
will correspond to $u=0$.}
$$
      uv = \left( 1-\frac{r}{a} \right) e^{r/a}  \ .   \eqno(2)
$$
The associated volume element is 
$$
   d^2V = \sqrt{-g} \, du dv = {2a^3\over r} e^{-r/a} du dv 
   \ . 
   \eqno(3)
$$

Now let $\Sigma$ be the ingoing null hypersurface defined by the
equation $v=v_0$, and let $(x,y)$ be a pair of sprinkled points
satisfying the following conditions:
$$
  \left\{ 
  \begin{array}{c}
    x\in J^{-}(\Sigma) \cap J^{-}(H) \\ 
    y\in J^{+}(\Sigma) \cap J^{+}(H) \\ 
    x \prec y \mbox{ a link }        \\	
    x \mbox{ maximal in } J^{-}(\Sigma) \cap J^{-}(H) \\ 
    y \mbox{ minimal in } J^{+}(H)
  \end{array}
  \right\}
  \eqno(4)
$$
(For a null $\Sigma$ in two dimensions, the fourth condition is actually
redundant, but it would be needed with a spacelike $\Sigma$.)  In order
that these conditions be fulfilled, no sprinkled point (other than $x$
or $y$) must fall into the shaded region depicted in Figure~1.  The
volume of this excluded region is readily evaluated and is given by
$$
     V = a^{2} + r_{xy}^{2} - r_{xx}^{2} - r_{yy}^{2}  \ ,
$$
where we have adopted the notation,
$$
     u_i v_j = \left(1-\frac{r_{ij}}{a}\right) e^{r_{ij}/a}  
     \ .
     \eqno(5)
$$
In analogy with equation (1),
the expected number of links satisfying our conditions is therefore
$$
  \bra{n}\ket
  \; =
  \left(2a^3\right)^{2}
  \int_{0}^{v_{0}} dv_{x}
  \int_{-\infty}^{0} du_{x}
  \int_{v_{0}}^{\infty} dv_{y}
  \int_{0}^{1/v_{y}} du_{y}
  \frac {e^{-r_{xx}/a - r_{yy}/a}} {r_{xx} r_{yy}} \; e^{-V}
$$ 

A change of integration variables 
from $(u_x,v_x,u_y,v_y)$ to $(r_{xx},r_{x0},r_{xy},r_{yy})$,
followed by the notational substitutions
$x=r_{xy}$, $y=r_{x0}$, $z=r_{xx}$,
now reduces $\bra{n}\ket$ to the form,\footnote%
{Here $r_{x0}$ is of course the radial variable corresponding via
 (5) to the product $u_xv_0$.  To avoid confusion, notice that the
 dummy integration variables $x$, $y$ and $z$ are real numbers entirely
 distinct from the sprinkled points $x$ and $y$.}
$$
  \bra n \ket \; = 4 \, I(a) \, J(a)  \ ,
$$
where
$$
  I(a) 
  =
  \int_{a}^{\infty }dx \frac{x}{x-a} e^{-x^{2}} 
  \int_{a}^{x}dy \frac{y}{y-a} 
  \int_{a}^{y} e^{z^{2}}dz
  \eqno(6)
$$
and
$$ 
  J(a) = e^{-a^{2}} \int_{0}^{a} e^{r_{yy}^2} dr_{yy} \ . 
  \eqno(7)
$$
Notice that $\bra{n}\ket$ does not depend on $v_0$, reflecting the
stationarity of the black hole.

% %% RDS I think this is nontrivial only for spacelike Sigma, but here,
% %% we're dealing with the null case.
% %
% In fact any dependence on $v_{0}$ (at least the leading contribution)
% would kill any hope for this calculation to produce any thing of
% physical significance, for stationary black holes.

Now, inasmuch as comparison with the Bekenstein-Hawking entropy is
meaningful only for macroscopic black holes, we might as well assume
that $a\gg{1}$, and in that regime, $I(a)$ can be shown [5]
to have the following asymptotic behavior:
$$
    I(a) = {\pi^2 \over 12} \; a + O\left({1 \over a}\right)  \ .
$$
On the other hand it is not difficult to see that
$$
      J(a) = \frac{1}{2a} + O\left( {1 \over a^3} \right)  \ .
$$
%%       e^{-a^{2}} \int_{0}^{a} e^{r_{yy}^{2}} dr_{yy} 
Putting everything together, we end up with
$$
    \bra{n}\ket\; = {\pi^2\over6} + O\left( 1/a \right)  \ .
    \eqno(8)
$$

Although our calculation has been carried out in two dimensions, a study
of the integrals $I(a)$ and $J(a)$ indicates that, were we to redo it in
four dimensions, the expected number of links would reduce to
essentially the same expression.  Indeed, the dominant contribution to the
integral $J(a)$ plainly comes from $r_{yy}\approx{a}$, but since
$r_{yy}$ is the radial coordinate $r$ of sprinkled point $y$, and since
$r=a$ is the horizon, this implies that $y$ resides near the horizon.
Similarly, the dominant contribution to the integral $I(a)$ comes from
$z\approx{a}$, which, since $z=r_{xx}$, implies in turn that sprinkled
point $x$ resides near the horizon as well.  Consequently our counting
can be said to be controlled by the near horizon geometry.  But in four
dimensions, this geometry is locally just the two dimensional one times
the Euclidean plane.  Thus, one would expect $\bra{n}\ket$ to be simply
proportional to the area of the horizon.  Moreover, from (8),
one would expect the coefficient of proportionality to be of order
unity, although there is of course no reason for it to be exactly
$\pi^2/6$.

It is interesting that part of what makes the near horizon pairs special
is the vanishing of the denominators in $I(a)$ when the dummy
integration variables $x$ and $y$ tend to $a$.  To the extent that it is
this divergence which makes the horizon such a strong source for the
links, we may be reminded of the analogous fact that the strong redshift
in the vicinity of the horizon allows modes of arbitrarily high (local)
frequency to contribute to the entanglement entropy without influencing
the energy as seen from infinity.  Notice also that the clustering of
$x$ and $y$ near the horizon is not simply a consequence of the
maximality and minimality conditions we imposed on them.  For instance,
pairs $(x,y)$ sitting arbitrarily close to the hypersurface $\Sigma$,
with $y$ arbitrarily close to the horizon, still do not contribute to the
leading term in $I(a)$ if $x$ is far from the horizon, namely with
coordinate $\left|u_{x}\right|\gg1$.

%% (RDS) is the above rewrite ok? 

\subsection{A black hole far from equilibrium}

% This is figure 2
\begin{figure}[t]
\epsfxsize=3.28truein
\centerline{\epsfbox{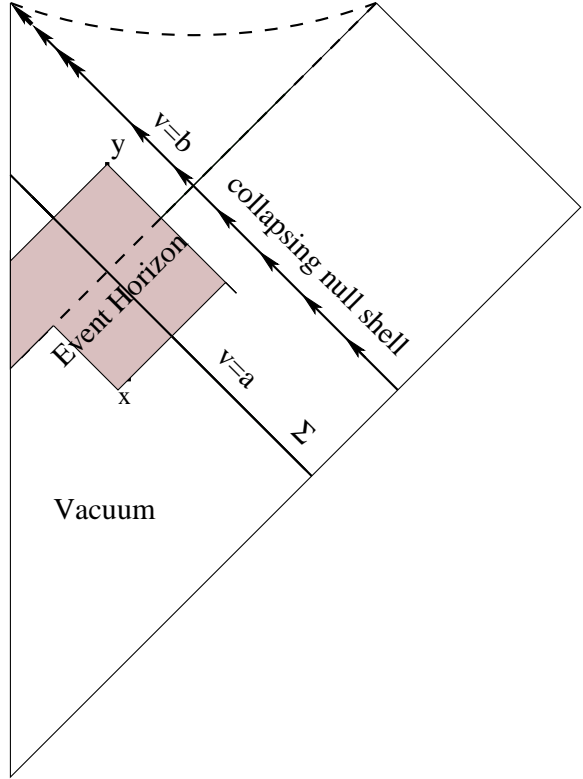}}
\caption{A non-stationary horizon and null hypersurface $\Sigma$}
\end{figure} 

Turning now to a case which, though still spherically symmetric, is very
far from equilibrium, let us consider a shell of null matter which
collapses to form a Schwarzschild black hole.
%
% (RDS) I removed the formula, because I couldn't sort out the
% correct coefficient.  Feel free to reinstate if you know it. 
% with stress tensor given by
% $$
%      T_{vv} = { M\delta(v-b) \over 4 \pi r^{2}}   \ ,
% $$
% the other components all being zero.  
%
The Penrose diagram for this spacetime\footnote%
{A fuller description of this spacetime may be found, for example,
 in [6].}
is shown in Figure~2.  Let the
shell sweep out the world sheet $v=b$ and let us choose for our
hypersurface $\Sigma$ a second ingoing null surface defined by $v=a$,
with $a<b$ so that $\Sigma$ lies wholly in the flat region.  Here $u$
and $v$ are null coordinates, chosen so that the horizon first forms at
$u=v=0$ and normalized for convenience such that
$$
       ds^2 = - 2 du dv + r^2 d\Omega^2 \ .
$$
Since our interest is again in macroscopic black holes, we will assume as
before that the horizon radius is large in units such that $\varrho_c=1$;
and to simplify matters further, we will also restrict ourselves to a
time well before the infalling matter arrives
(as judged in the center of mass frame).
We thus have the double inequality, $b{\gg}a{\gg}1$.  
Once again, we will perform the calculation for the two dimensional
radial section rather than the full four dimensional spacetime.

Now since we are assuming that the infalling matter is far to the future
of the hypersurface $\Sigma$, points $y$ sprinkled into that region
should not contribute significantly when our minimality and link
conditions are taken into account.  For this reason, we shall, for
convenience, restrict our counting to pairs $(x,y)$ with $v_y<b$.
Imposing, then, the same 
conditions (4) introduced above, we
obtain for the expected number of causal links
$$
  \bra n \ket  \;
  =
  \int_{a}^{b} dv_{y} 
  \int_{0}^{v_{y}} du_{y} 
  \int_{-\infty}^{0} du_{x}
  \int_{0}^{a} dv_{x}  \,
  e^{-V}
$$
where $V=u_{y}v_{y}-u_{x} (v_{y}-v_{x}) - u_{y}^{2}/2$.

\smallskip

It is not difficult to derive the leading behavior of this integral for
large $a$, and here we quote only the final result:
$$
  \bra n \ket  \;
   =
   {\pi^2 \over 6} - l  \left({a \over b}\right) + O(1/a)  \ ,
$$
where $l(x)\ideq\sum_{k=1}^\infty{x}^k/k^2$, a convergent series that
vanishes in the limit $x\to{}0$.  
Since we have assumed that $a{\ll}b$, we can write this more simply as
$$
  \bra{n}\ket\; = {\pi^2 \over 6} + O(a/b) + O(1/a) \ .   \eqno(9)
$$
Notice that the presence of a negative contribution like $-l(a/b)$ was
to be expected, since we have omitted to count links that extend past
the shell into the Schwarzschild region.  For $\Sigma$ near to the
shell, one obviously should not neglect such links, and our counting is
incomplete. 

Two features of the result (9) are especially noteworthy.  The
first is its independence of the value of $a$.  As with the equilibrium
black hole above, this indicates that the analogous four dimensional
computation would produce (at leading order) an answer proportional to
the horizon area.  What is then even more striking is the occurrence of
the same numerical coefficient ${\pi^2}/{6}$ in both (8) and
(9).  This agreement furnishes a nontrivial consistency check
of the suggestion that one can attribute the horizon entropy to the
``causal links'' crossing it.

Now what can one say about the case where the hypersurface $\Sigma$ is
spacelike?  In two dimensions it can be shown that $\bra{n}\ket$ is
again finite and independent of $a$ to leading order.  Although we have
not carried the calculation far enough to verify explicitly that one
obtains for it the same numerical answer $\pi^2/6$,
one can make it plausible on general grounds that this would have to
happen.  The point is that (in the flat case) the definition of
$\bra{n}\ket$ is manifestly Lorentz invariant, whence any spacelike
plane (or in this case line) $\Sigma$ must give the same answer as any
other related to it by a boost.  But in the limit of tilting, a
spacelike line becomes null, and by continuity the corresponding
$\bra{n}\ket$ should go over to $\pi^2/6$ in this limit.  Now observe
that a suitable boost transformation will convert any nearly null line
$\Sigma$ into one which is ``purely spacelike'' (and with a larger value
of $a$).  This gives a good reason to expect that both null and
spacelike $\Sigma$ must yield the same result.  Observe also, that a
similar argument can be made for the Schwarzschild case, using the
time-translation Killing vector instead of the boost Killing vector.

In four dimensions, the calculation of $\bra{n}\ket$ needs a much more
elaborated technique, both for null and spacelike hypersurfaces.  The
calculation of the volumes needed to insure the conditions
(4) is lengthy, and it turns out that one has to distinguish
many cases depending on the relative positions of the
linked points $x$ and $y$, each case making its own contribution to
$\bra{n}\ket$.  Fortunately, only a few of these contributions survive
for macroscopic black holes ($a\gg1$), and it should be
possible to evaluate them all with sufficient effort.  Here we give only
the final result for one such contribution, referring the reader to
reference [5] for further detail:
$$
   <n>_{1} = \frac{\pi^3a^2}{16} \left( c + O\left( 1/a \right) \right)
$$
where
$$
   c =
   \int_{0}^{\infty} dx 
   \int_{0}^{x} dy
   \left(x-y\right)^{4} 
   e^{ -\frac{\pi}{3} \left(x^{4}+y^{4}\right) }
   \approx 0.0419
   %% c = 0.0418608...
$$
\smallskip
As indicated above, it is not difficult to convince oneself on the basis
of our two dimensional experiences that the number of links in four
dimensions must turn out to be proportional to the area of the horizon,
or more precisely, to the area of the two-surface $S=H\cap\Sigma$.  To
recall the reasoning: The surfaces $H$ and $\Sigma$ will look locally
like their two dimensional analogs extended trivially by a portion of
$\Reals^2$, but since, as we saw, the main contribution to $\bra{n}\ket$
in two dimensions came from pairs just straddling $H\cap\Sigma$, and
since locally $\Sigma$ will also look flat (like our two dimensional
$\Sigma$ was), and since (as we argued) all flat $\Sigma$ (null or
spacelike) give the same (finite) answer in two dimensions, so in four
dimensions the density of links per unit surface area of $S$ will be
constant, that is to say, their total number will be proportional to the
area of $S$, modulo subleading corrections.  Moreover, the same should
hold for arbitrary hypersurfaces $\Sigma$ and arbitrarily curved
horizons $H$, as long as neither is so badly distorted as to exhibit
significant curvature in the vicinity of a horizon point.  Accepting all
this, we can anticipate the general formula for four dimensions:
$$
  \bra{n}\ket\; 
  =
  \gamma
  { A \left(H\cap\Sigma\right) \over l_c^2 }
  \left( 
    1 + O \left({l_c \over \sqrt{A\left(H\cap\Sigma\right)}}\right) 
  \right)
  \ ,
$$
where $A(H\cap\Sigma)$ is the area of the 2-surface in which the horizon
meets $\Sigma$, $l_c=\varrho_c^{-1/4}$ is the fundamental causal set
length, and $\gamma$ is a number of order unity.  For macroscopic black
holes we can safely neglect the second term and conclude that the number
of links will just be proportional to the area of the horizon in causal
set units, with a coefficient of order unity.  From this we can infer
that, if the entropy really does measure the number of causal links,
then $l_c$ must be of Planckian order, as was anticipated a long time
ago.

\section{On the minimality and maximality conditions}

% This is figure 3
\begin{figure}[t]
\epsfxsize=3.28 truein
\centerline{\epsfbox{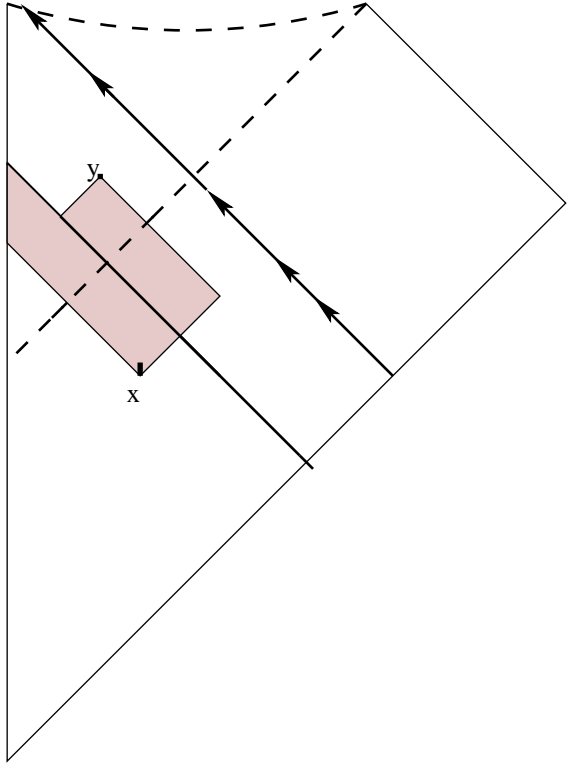}}
\caption{A second variation on the ``max/min'' theme}
\end{figure}

Now we return to the ``max/min'' conditions we introduced in
Section~2, in order to prevent the double counting of causal links to
which we attributed the initially divergent character of our integral
for $\bra{n}\ket$.  In (4), these conditions are the last
two in the list.  Other possibilities exist, however, and we know of
nothing particularly sacred about the conditions used in
(4), which we selected partly with an eye to the simplicity
(for evaluation) of the resulting integral.  One must be careful not to
use something like ``$y$ minimal in $J^+(\Sigma)$'', which would drive
$\bra{n}\ket$ to zero in the limit of null $\Sigma$, but this does not
rule out, for example, a condition like ``$x$ maximal in
$J^-(\Sigma)$''.

Fortunately, the finiteness of the answer --- and even its exact
numerical value --- seems to be insensitive to variations in the max/min
conditions.  Consider, for example, repeating the calculation of
Section~2.2 with the different set of conditions illustrated in Figure~3.
%
%% (RDS) Djamel, we _are_ talking about the flat calc here, nes pas? 
%
(We have weakened the fifth condition to 
``$y$ minimal in $J^+(H)\cap{}J^+(\Sigma)$'' and strengthened the fourth to 
``$x$ maximal in $J^-(\Sigma)$''.)  
With this alternative set of conditions, the integral for $\bra{n}\ket$
is modified (because $V$ is modified), but it can be shown
[5] to have the same asymptotic behavior as before, namely
$$
     \bra{n}\ket\; = \frac{\pi^2}{6} + O(1/a)  \ .
$$
Thus, in this case at least, we obtain exactly the same numerical answer
as in Section 2.

Another feature that our counting must have if it is to yield the
horizon area is that, within reason, the expected number of links should
depend only on the intersection $H\cap\Sigma$, and not on how the
surface $\Sigma$ is prolonged outside or (especially) inside the
horizon $H$.  For example one should get the same answer for both of
the continuations shown in Figure 4. (The case where the difference is
confined to the interior black hole region is of particular significance
for the entanglement interpretation of horizon entropy, since such a
difference cannot, by definition, influence the effective density
operator for the external portion of $\Sigma$ (at least to the extent
that unitary quantum field theory is a good guide).)  From this point of
view, those max/min conditions are most satisfactory that depend least
on conditions outside the black hole.  In this sense, the condition used
in Section~2 that $y$ be minimal in $J^{+}(H)$ has an advantage over the
alternative, ``dual'' condition that $x$ be maximal in $J^{-}(\Sigma)$;
for the former, at least in the case of null $\Sigma$, refers only to
the interior region.

% This is figure 4
\begin{figure}[t]
\epsfxsize=3.28 truein
\centerline{\epsfbox{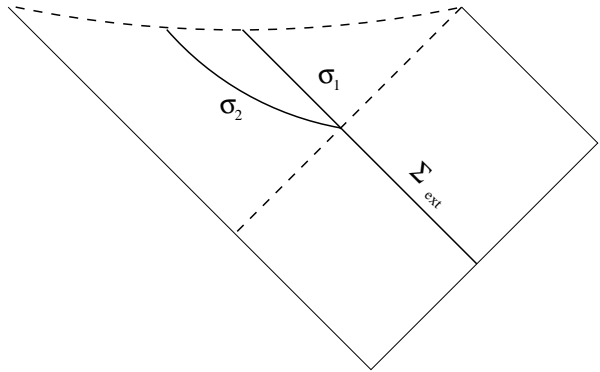}}
\caption{Two continuations of a hypersurface to the interior region.}
\end{figure}

\section{Summary and Discussion}

In Sections 2 and 3 we have reported some calculations, in the context
of causal set theory, of the expected number of irreducible causal
relations (links) that cross the horizon $H$ of a black hole in
proximity to a specified spacelike or null hypersurface $\Sigma$, as
determined by the satisfaction of certain ``max/min'' conditions.
Limiting ourselves to the case of spherical symmetry, we considered both
equilibrium and nonequilibrium examples of macroscopic black holes in
both 3+1 and 1+1 dimensions, together with both null and spacelike
hypersurfaces.  We also considered variants of the max/min conditions.
In all these cases one obtains finite answers, but we computed exact
numbers only for null $\Sigma$, and only for the two dimensional
reductions of the corresponding four dimensional black holes.  The
expected number of links was always $\pi^2/6$.  Moreover, we saw that the
bulk of the links always resided in close proximity to the horizon,
meaning that the result was being controlled by the near horizon
geometry.  From this we inferred the likelihood of a universal
relationship in four dimensions, with the number of links being
proportional to the horizon area, modulo corrections down by a factor of
$l_c/R$ where $l_c$ is the fundamental discreteness length and $R$ the
black hole size.\footnote%
{Interestingly, these corrections are -- in 4 dimensions -- comparable
 in order of magnitude to the inherent $\sqrt{n}$ fluctuations that one
 would expect in $n$ itself purely for statistical reasons.}

What seems significant about these results is not so much the
proportionality to horizon area {\it per se}.  One might have expected
as much.  However the coefficient of proportionality might have turned
out to be either infinite or zero in the limit of large black hole
radius (as in fact it does if one omits the max/min conditions
introduced to prevent ``double counting'' of links).  Moreover, for the
nonequilibrium horizon, the coefficient might have varied with time or
it might have differed from its equilibrium (Schwarzschild) value.  In
the event, none of these things occurred, at least in the cases checked.
Rather we found a universal answer which took the same value in all
cases where we succeeded in evaluating it exactly.  The agreement
between the equilibrium and nonequilibrium cases seems especially
noteworthy, inasmuch as this is the first time, to our knowledge, that
such an entropy has been computed for a non-stationary horizon.

The weakness of our result, of course, is that it remains at a purely
kinematic level: we believe to have found something like the number of
``horizon atoms'', whose multiplicity is the ultimate source of black
hole entropy, but this belief cannot be substantiated or refuted before
we possess a fully quantum dynamics for causal sets.  Short of this,
many interesting extensions and cross-checks of our conclusions can
still be pursued, however.

Of greatest immediacy is the need to carry out a full calculation in 
four-dimensions.  Not only would this provide an important test
of our reduction to two dimensions, but it would furnish the correction
to the two-dimensional value of $\pi^2/6$, thereby laying the basis for a
future determination of the fundamental length $l_c$.  (Comparison of
the known entropy with a calculation from first principles is probably
the most reliable way to get a handle on the basic parameters of any
quantum gravity theory, as entropy, being an absolute number, should not
be subject to ``renormalization''.)

Completing the evaluation of the two dimensional integrals for spacelike
$\Sigma$ is also desirable in order to decide whether they indeed give
the same result as for null $\Sigma$.\footnote%
{In this connection, we note that the case of null $\Sigma$ is of
 particular interest for the ``Generalized Second Law'' of entropy
 increase.  It is difficult to imagine proving this law --- or even
 formulating it --- without being able to specify in a well defined
 manner the hypersurface $\Sigma$ to which the entropy is being
 referred.  Within a semiclassical spacetime with its fixed metric, this
 is not a problem, but the semiclassical framework is overly
 restrictive, since it cannot accommodate, for example, such a mundane
 entropy as that due to the spread in position of the individual members
 of ``gas'' of black holes.  Fortunately, recourse to a semiclassical
 spacetime is unnecessary in the case of a null hypersurface, since then
 one can specify $\Sigma$ by ``anchoring it to the environment'' (say to
 the walls of the proverbial thermodynamic box), and with this
 accomplished, one can envisage proving the second law as sketched in
 [7].  But no similarly robust technique seems
 available in the case of a spacelike $\Sigma$.}
It would also be good to explore further the extent to which the results
we have obtained depend on the details of the max/min conditions chosen
(or indeed, whether other conditions, not of the max/min type might possibly
offer a better solution to the ``double counting'' problem).  The
generalization to rotating and deformed black holes is another obvious
direction for further work.

With respect to the causal set, there can of course be no horizon as
such, only a division of the elements into those which can and cannot
communicate with distant regions.  The closest one can come to the
horizon as a null surface is probably the collection of linked pairs we
have counted in this paper.\footnote%
{Another possibility might be the minimal layer $L$ of the subcauset
 corresponding to the interior region of the black hole.  However, $L$
 is by definition an antichain and therefore more akin to a spacelike
 surface than a horizon, which, though not everywhere null, is ruled by
 null geodesics.  More importantly, as one moves along $H$ toward the
 future, the elements of $L$ probably become sparse too rapidly to mark
 out $H$ correctly.  This difficulty is even clearer for the dually
 defined set $L'$ of {\it maximal} elements of the {\it exterior}
 region, which probably is empty!}
But these correspond to a ``thickened hypersurface'' in the continuum.
It would be interesting to compute the amount of this ``kinematical
thickening'', especially as there are hints from a very different
direction of a pronounced ``dynamical thickening'' of the horizon
(possibly of order $a^{1/3}$) resulting from the influence of quantum
fluctuations in fields propagating near $H$ [8].

A further direction for generalization would be the substitution of some
different structure for the links we have considered in this paper.  One
such possibility might be a ``triad'' of elements, say $x$, $y$ and $z$
with $x$ and $y$ to the past of $\Sigma$ and $z$ to its future, and with
$x$ inside the black hole and $y$ outside it.  The requirement that the
triad be ``small'' in a suitable sense might then be able to replace our
max/min conditions.  In the same spirit one could consider inverted
triads or even ``diamonds'' containing both types of triad
simultaneously.  There is however some suggestion that triads of the
first type are naturally related to the kind of correlation responsible
for entanglement entropy in a quantum field theory framework
[9].

A final remark concerns the finiteness of our integrals in two spacetime
dimensions.  Although this was necessary in order that the four
dimensional result scale correctly with area, it could nonetheless seem
surprising that the counting of two dimensional links remains finite
even in the continuum limit where the fundamental length is sent to
zero.  In this sense, the replacement of continuous spacetime by a
causal set could appear in two dimensions as more of a regularization
device than something fundamental.  We do not know whether this has any
deeper meaning, or whether it might be related to some of the other
special properties that both quantum field theory and quantum gravity
possess in two dimensions (cf. [10]).

In concluding, we would like to dedicate this article to our friend and
colleague, Jacob Bekenstein.  Not only does Jacob's work lie at the
origin of our understanding of black hole entropy and the ``generalized
second law'', but it also raised explicitly the theme of missing
information which forms the backdrop to, and inspiration for, the work
reported herein.

The work of RDS was partly supported by NSF grant PHY-0098488, by a
grant from the Office of Research and Computing of Syracuse University,
and by an EPSRC Senior Fellowship at Queen Mary, University of London.
The work of DjD was partly supported by grant number ERBFMRXCT 960090.
RDS would like to thank Goodenough College, London for its hospitality
during the writing of this paper.
DjD would like to thank the General Relativity group of Syracuse
University and the High Energy group of ICTP for their kind hospitality
during different stages of the work.

\ReferencesBegin

\reference [1]
J.D.~Bekenstein,                           %  Jacob D. Bekenstein (Hebrew U.)
 ``Do we understand black hole entropy?'',
   in {\it The Seventh Marcel Grossmann Meeting on Recent Developments in
    Theoretical and Experimental General Relativity, Gravitation and
    Relativistic Field Theories},
    proceedings of the MG7 meeting, held Stanford, July 24--30, 1994,
    edited by R.T.~Jantzen, G.~Mac~Keiser and R.~Ruffini
   (World Scientific 1996)
   \eprint{gr-qc/9409015}

% \reference [R::horizon-atoms] %% NEEDED
%
% Some reference speculating on horizon atoms in causet theory.
%
%    % ref $\left[ 1\right] $ of djamel

\reference [2]
L.~Bombelli, J.~Lee, D.~Meyer and R.D.~Sorkin, ``Spacetime as a Causal Set'', 
\journaldata{Phys. Rev. Lett.}{59}{521-524}{1987};
\linebreak
%
% ``Spacetime and Causal Sets'', 
%      in J.C. D'Olivo, E. Nahmad-Achar, M. Rosenbaum, M.P. Ryan, 
%               L.F. Urrutia and F. Zertuche (eds.), 
%     {\it Relativity and Gravitation:  Classical and Quantum} 
%     (Proceedings of the {\it SILARG VII Conference}, 
%       held Cocoyoc, Mexico, December, 1990), 
%     pages 150-173
%     (World Scientific, Singapore, 1991);
% \linebreak
%
David Porter Rideout and Rafael Dolnick Sorkin,
``A Classical Sequential Growth Dynamics for Caus\-al Sets'',
 \journaldata{Phys. Rev. D}{61}{024002}{2000}
 \eprint{gr-qc/9904062}

\reference [3]
L.~Bombelli, ``Statistical Lorentzian geometry and the closeness of
 Lor\-entz\-ian manifolds'', 
\journaldata{J. Math. Phys.}{41}{6944-6958}{2000} 
\eprint{gr-qc/0002053}

\reference [4]
G.~Brightwell and R.~Gregory, ``The Structure of Random Discrete Spacetime'',
    \journaldata{Phys. Rev. Lett.}{66}{260-263}{1991}  

\reference [5]
 Djamel Dou,			
  {\it Causal Sets, a Possible Interpretation for the Black Hole
  Entropy, and Related Topics}, 
  Ph.~D. thesis (SISSA, Trieste, 1999)
  \eprint{gr-qc/0106024}

\reference [6]
Rafael D.~Sorkin and Daniel Sudarsky,
``Large Fluctuations in the Horizon Area and What They Can Tell Us About
   Entropy and Quantum Gravity'',
  \journaldata {Class. Quant. Grav.}{16}{3835-3857}{1999}
  \eprint {gr-qc/9902051}		

\reference [7]
 R.D.~Sorkin, 
``Toward an Explanation of Entropy Increase 
  in the Presence of Quantum Black Holes'',
  \journaldata {Phys. Rev. Lett.} {56} {1885-1888} {1986};
\linebreak
  R.D.~Sorkin,
``The Statistical Mechanics of Black Hole Thermodynamics'',
  in R.M. Wald (ed.) {\it Black Holes and Relativistic Stars}, 
  (U. of Chicago Press, 1998), pp. 177-194
  \eprint{gr-qc/9705006}

\reference [8] % ref on wrinkled horizon 
R.D.~Sorkin,
``How Wrinkled is the Surface of a Black Hole?'',
  in David Wiltshire (ed.), 
  {\it Proceedings of the First Australasian Conference on General
       Relativity and Gravitation}, 
  held February 1996, Adelaide, Australia, pp. 163-174
  (University of Adelaide, 1996)
  \eprint{gr-qc/9701056}

\reference [9]
  R.D.~Sorkin,
``Quantum Measure Theory and its Interpretation'', 
  in
   {\it Quantum Classical Correspondence:  Proceedings of the $4^{\rm th}$ 
    Drexel Symposium on Quantum Nonintegrability},
     held Philadelphia, September 8-11, 1994,
    edited by D.H.~Feng and B-L~Hu, 
    pages 229--251
    (International Press, Cambridge Mass. 1997)
    \eprint{gr-qc/9507057}

\reference [10]
Thomas M. Fiola, John Preskill, Andrew Strominger, and Sandip P. Trivedi,
``Black hole thermodynamics and information loss in two dimensions'',
\eprint{hep-th/9403137}
\journaldata{Phys.Rev. D}{50}{3987-4014}{1994}

\end{document}

%% file: mathmacros.tex
\font\openface=msbm10 at10pt
\def\Reals         {{\hbox{\openface R}}}

\def\past   {\mathop {\rm past     }\nolimits}
\def\future {\mathop {\rm future   }\nolimits}

 \def\dal{\displaystyle{{\hbox to 0pt{$\sqcup$\hss}}\sqcap}}

\def\lto{\mathop
        {\hbox{${\lower3.8pt\hbox{$<$}}\atop{\raise0.2pt\hbox{$\sim$}}$}}}
\def\gto{\mathop
        {\hbox{${\lower3.8pt\hbox{$>$}}\atop{\raise0.2pt\hbox{$\sim$}}$}}}
\def\braces#1{ \{ #1 \} }

\def\bra{<\!}			% These seem more useful than the two
\def\ket{\!>}			% commented out ones just below
\def\to{\mathop\rightarrow}	% symbol used eg in f:A-->B

\def\ideq{\equiv}		% triple equal sign

\def\SetOf#1#2{\left\{ #1  \,|\, #2 \right\} }

\def\interior #1 {  \buildrel\circ\over  #1}     % seems to work
\def\basisvector#1#2#3{
 \lower6pt\hbox{
  ${\buildrel{\displaystyle #1}\over{\scriptscriptstyle(#2)}}$}^#3}